\documentclass[prb,reprint,showpacs,floatfix,epsf,epsfig,twocolumn]{revtex4} 
\usepackage{epsf}
\usepackage{epsfig}
\usepackage{epstopdf}
\usepackage{graphicx}

\begin{document}

\title{Magnetic Interactions and Electronic Structure of Pt$_{2}$Mn$_{1-x}$Y$_{x}$Ga (Y = Cr and Fe) system : An ab-initio calculation}

\author{Tufan Roy$^{1}$, Aparna Chakrabarti$^{1,2}$\footnote{Electronic mail: aparnachakrabarti@gmail.com}}
\affiliation{$^{1}$ HBNI, Raja Ramanna Centre for Advanced Technology, Indore - 452013, India}
\affiliation{$^{2}$ ISUD,Raja Ramanna Centre for Advanced Technology, Indore
- 452013, India}

\begin{abstract}
First principles density functional theory based calculations have been 
carried out to predict the effects of Mn replacement by Fe and 
Cr on electronic as well as magnetic properties of  Pt$_{2}$MnGa
 as well as Ni$_{2}$MnGa. All the materials studied here are predicted to have conventional
   Heusler alloy structure in their ground state and they are found to be  electronically stable on the basis of their respective  formation energy. The replacement
    of Mn by Fe leads to a ferromagnetic ground state whereas in
     case of Mn replacement by Cr an {\it intra-sublattice} 
     anti-ferromagnetic configuration has been observed to have lower energy. 
We study the magnetic exchange interaction between the atoms 
for the materials with ferromagnetic and anti-ferromagnetic configurations to show the effects of Fe and Cr substitution at Mn site on the magnetic interactions of these systems. Detailed analysis of electronic structure in terms of density of states has been carried out to study the effect of substitution.
\end{abstract}


 
\maketitle


\section{Introduction}

Ni$_{2}$MnGa is one of the most studied magnetic shape memory alloys (MSMA). It shows magnetic field induced strain (MFIS) and large magnetoresistance effect (MRE), which make Ni-Mn-Ga a suitable candidate for practical application as actuators and sensors.\cite{PhilMag-webster-1984,apl-cbiswas-2005,apl-sozinov-2002}. The materials which are known as MSMA are having two different crystal structures, a high temperature phase with high crystal symmetry (cubic) known as austenitic phase and another one is the low temperature phase of lower symmetry (e.g. tetragonal, orthorhombic) named as martensitic phase. The temperature at which the structural transition occurs is known as martensitic transition temperature(T$_{M}$). For Ni$_{2}$MnGa, T$_{M}$ is reported to be 210 K.  Besides MFIS and MRE, the MSMAs are well known to show magnetocaloric effect (MCE)\cite{apl-hu-2000} and inverse magnetocaloric effect\cite{nat-mater-krenke-2005, nature-kainuma-2006}, which make them potential material of choice for use in echo-friendly refrigerator. 

However, a major drawback of Ni$_{2}$MnGa is its low (less than room temperature) martensitic transition temperature and its brittleness. Therefore, the current challenge for the researchers is to find new materials which possess better mechanical property and preferably higher T$_{M}$ and T$_{c}$ (Curie temperature) compared to Ni$_{2}$MnGa. It is reported in literature that both T$_{M}$ and T$_{c}$  are highly composition dependent. Experimentally it has been observed that partial Cu substitution at the Mn site of Ni$_{2}$MnGa raises the martensitic transition temperature\cite{prb-sroy-2009}. In our previous work\cite{prb-achakrabarti-2013} based on first principles calculations we have also shown that Cu substitution at Mn or Ga site improves the martensitic transition temperature whereas the Cu substitution at Ni site stabilizes the austenite phase more.

 Recently, on the basis of first principles calculations, 14\% MFIS has been reported for Ni$_{1.75}$Pt$_{0.25}$MnGa \cite{apl-mario-2011}. It has also been predicted that systematic as well as full substitution at Ni site by Pt, improves the mechanical property as well as martensitic transition temperature\cite{jalcom-troy-2015, jmmm-troy-2016}. In literature it is reported that Mn replacement by Fe in  Ni$_{2}$MnGa with suitable amount of substitution pushes T$_{M}$ close to room temperature\cite{apl-oikawa-2002,prb-zhang-2008}. Keeping the goal in mind for searching for better Heusler alloys with higher T$_{M}$ and  T$_{c}$ as well as lower inherent crystalline brittleness (ICB), we probed the replacement of Mn by Cr and Fe in our earlier work\cite{jmmm-troy-2016} and we have shown that full replacement of Mn by Fe decreases the ICB in both cases of Ni$_{2}$MnGa and Pt$_{2}$MnGa. There we have also reported significant increase of T$_{M}$ with respect to parent materials. Now it will be interesting to probe how the electronic density of states as well as magnetic exchange interaction are modified in case of Mn substitution by Fe and Cr. Therefore, here, on the basis of first principles calculations, we study systematically the effect of Fe and Cr substitution at the Mn site on their electronic properties, as well as on the T$_{M}$ and T$_{c}$ for materials Pt$_{2}$MnGa as well as Ni$_{2}$MnGa. Using spin polarized fully relativistic Korringa-Kohn-Rostoker (SPRKKR) method, we also study the Heisenberg magnetic exchange coupling parameters to understand the effect of substitution on the magnetic interaction in these materials. 

\section{Method}
For the first principles calculation, geometry optimization of all the materials studied here have been performed using  Viena Ab initio Simulation Package(VASP) \cite{prb-kreese-1996} along with projected augmented wave method (PAW)\cite{prb-blochl-1994}. Generalized gradient approximation (GGA) over local density approximation (LDA) of Perdew, Burke and Ernzerhof has been used for exchange correlation functional\cite{prb-pbe-1996}. We use energy cutoff for plane wave 500\,eV and convergence has been tested. The final energies have been calculated with a $k$-mesh of 15$\times$15$\times$15 for the cubic case. The energy and 
force tolerance used were 10 $\mu$eV and 10 meV/\AA, respectively. 

We have performed relativistic spin polarized all-electron calculation on the optimized geometry to understand magnetic as well as electronic properties, that is density of states (DOS). All the calculations have been carried out using full potential linearized augmented plane wave (FPLAPW) program\cite{wien-pblaha-2002}, with the generalized gradient approximation for the exchange correlation functional\cite{prb-pbe-1996}. Brillouin zone (BZ) integration has been carried out using the tetrahedron method with Bloechl corrections\cite{wien-pblaha-2002} to obtain the electronic properties. An energy cut-off for the plane wave expansion of about 16 Ry is used and the charge density cut-off used here is G$_{max}$=14. For cubic phase, the number of k points for the self-consistent field (SCF)
cycles in the reducible (irreducible) BZ is about 8000 (220).The SCF calculations have been performed with the tolerance for energy convergence 0.1 mRy per atom and charge convergence is set to 0.0001.

To probe the details of magnetic interaction between the atoms, the calculation of Heisenberg exchange coupling constant J$_{ij}$ has been carried out. SPRKKR programme, provided by Ebert \textit{et al }\cite{rep-ebert-2011} has been used to calculate J$_{ij}$ within a real space approach\cite{Liechtenstein-jmmm-1987}.The exchange and correlation term has been incorporated within the GGA framework\cite{prb-pbe-1996}. For SCF cycles number of k points has been taken as 1000. The angular momentum expansion up to l$_{max}$=3 has been taken for each atom.

\section{Results and Discussions}
\subsection{Electronic property}
From the total energy calculation it has been found that, for all these materials parent as well as substituted ones, conventional Heusler alloy structure is energetically lower compared to their inverse Heusler alloy structure. We have analysed the variation of formation energy for all the materials as a function of Fe or Cr substitution in our previous work\cite{jmmm-troy-2016}. There we have also shown from the variation of energy as a function of tetragonal distortion and the observation of conservation of volume in both cubic and tetragonal phases, that all the materials in their ferromagnetic (FM) phase are likely to show tetragonal distortion leading to the possibility of a martensitic transition. In this work, we are interested to see the electronic density of states of the cubic austenite phase of the substituted alloys in detail.

\begin{figure}[ht]
\begin{center}
\includegraphics[width=0.6\columnwidth]{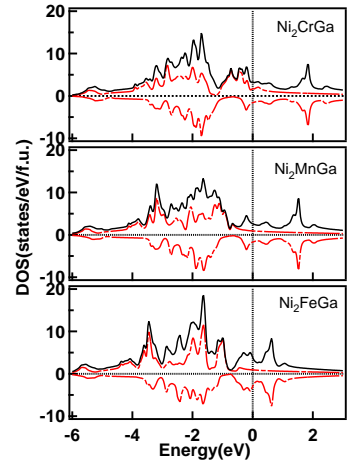}
\caption{Total DOS and Spin polarized DOS for Ni-based materials; solid line represents the total DOS and the majority and minority DOS are shown by dash-dotted lines}
\label{fig:appbimg}
\end{center}

\end{figure} 

First we discuss the DOS of the Ni$_{2}$YGa (Y = Cr, Mn, Fe) systems and we compare these results with the Pt$_{2}$YGa systems. Figure 1 shows the total DOS of all Ni-based materials along with spin polarized DOS. From the total DOS we can see that for Ni$_{2}$MnGa there is a double peak structure very close to Fermi level. The implication of this peak has been discussed in detail in our previous work and references therein\cite{jalcom-troy-2015}. In case of Ni$_{2}$MnGa, Barman \textit{et al} have shown that the peak in the minority DOS very close to Fermi level plays a crucial role in favouring martensitic transition\cite{prb-srbarman-2005}. Here from Figure 1, it is clear that all the Ni-based materials are having a prominent Ni d-electron derived peak very close to (also below) the Fermi level. While there, for Fe and Mn cases, the minority spin makes the major contribution, in case of Cr, both the spin densities of states show similar intensity. We note that it is expected from their respective DOS that tetragonal distortion may lower the energy of these systems compared to their cubic structure, which can be very well explained by band Jann-Teller distortion mechanism as has been observed in case of Ni$_{2}$MnGa in the literature\cite{prb-srbarman-2005}. 

\begin{figure}[ht]
\begin{center}
\includegraphics[width=0.6\columnwidth]{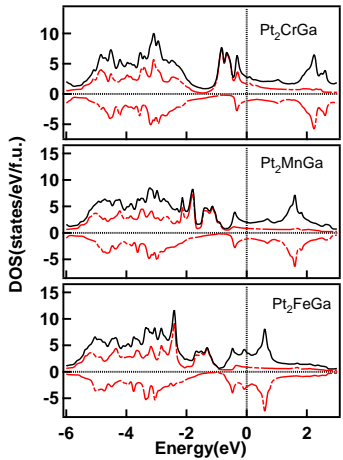}
\caption{Total DOS and Spin polarized DOS for Pt-based materials; solid line represents the total DOS and the majority and minority DOS are shown by dash-dotted lines}
\label{fig:appbimg}
\end{center}
\end{figure}

Similarly for Pt-based systems also, from Figure 2, we can find such peaks in the DOS very close to Fermi level; this observation  favors the possibility of martensitic transition. It is interesting to observe that the contribution of the majority and minority spin in this feature of the DOS close to Fermi level is similar in both the Pt and Ni-based systems.

\begin{figure}[ht]
\begin{center}
\includegraphics[width=0.6\columnwidth]{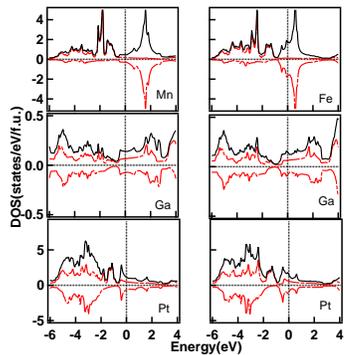}
\caption{Partial spin polarized DOS for Pt-based materials}
\label{fig:appbimg}
\end{center}
\end{figure}

To understand the electronic structure of the Pt-based materials in greater details, the partial DOS are plotted in Figure 3. From the partial DOS of the Pt$_{2}$MnGa and Pt$_{2}$FeGa it is seen that
the separation of majority and minority DOS for Mn and Fe, respectively, is clear and complete. What we mean by this is the following. We observe that for both the materials
Pt$_{2}$YGa (Y = Mn, Fe), the occupied DOS of the Y atom is dominated by the majority spin and the respective unoccupied DOS is dominated by the minority spin. 
For Pt$_{2}$MnGa majority DOS of the Mn atom is centred around -1.242 eV whereas the respective minority DOS is centred around 1.588 eV. For Pt$_{2}$FeGa
majority DOS of the Fe atom is centred around  -1.505 eV and respective minority DOS is centred around 0.617 eV. This gives the idea of a large exchange splitting in case of Y atom in both Pt$_{2}$MnGa and Pt$_{2}$FeGa, which is evidently more for the Mn atom. This large exchange splitting of Y atom for both the compounds leads to 
large partial magnetic moments on the Y atom (3.671$\mu$$_{B}$ for Mn in Pt$_{2}$MnGa, 3.025$\mu$$_{B}$ for Fe in Pt$_{2}$FeGa ) in both the materials\cite{jmmm-troy-2016}.
On the contrary, for the Pt atom, in both cases (Figure 3), the majority and minority DOS are very similar in peak position and intensity in
the occupied region, which leads to a much lower partial moment of the Pt atom (0.138$\mu$$_{B}$ in Pt$_{2}$MnGa, 0.088$\mu$$_{B}$ in Pt$_{2}$FeGa)\cite{jmmm-troy-2016}.
For both  Pt$_{2}$MnGa and Pt$_{2}$FeGa, there is significant hybridization between Ga 4p and 
outermost d electrons of the Pt atom and a double peak structure around Fermi level is observed which plays a crucial role
in the stability of these materials. This observation is similar to our earlier work.\cite{jalcom-troy-2015}

\subsection{Magnetic Interaction} 
In the literature it is well  established that Ni$_{2}$MnGa possesses ferromagnetic ground state configuration. For Pt$_{2}$MnGa, some studies suggest \cite{apl-mario-2011,thesis-antje,jalcom-troy-2015} that it is having ferromagnetic ground state. Therefore, we start the calculations for the substituted materials where Mn is replaced by Fe and Cr, by considering them as ferromagnetic in nature. The total magnetic moments of these materials along with partial moments are reported in detail in our previous work\cite{jmmm-troy-2016}. We observed that for Pt-based systems, more localization of Mn-magnetic moment is expected compared to the Ni-based systems because of the larger Mn-Mn separation in the former.
To get detailed insight of the magnetic interaction, in this work, we study the Heisenberg exchange coupling constants for all the materials in their cubic austenite phase. These coupling parameters give indication about the nature of the predominant magnetic interactions between the different atoms present in the materials.

\begin{figure}[htp]
\begin{center}
\includegraphics[width=0.6\columnwidth]{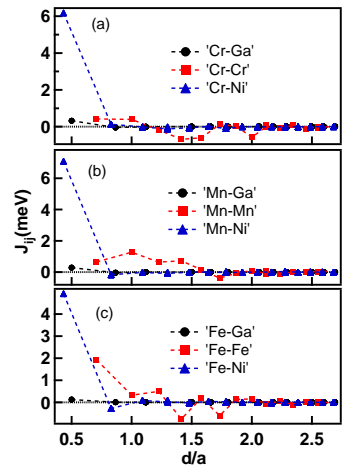}
\caption{Heisenberg exchange parameters J$_{ij}$ of Y atom (Y = Cr, Mn, Fe) with all the atoms as a function of normalized distance $d$$/$$a$ where $a$ is the lattice constant of the austenitic phase for (a)Ni$_{2}$CrGa (b) Ni$_{2}$MnGa and (c) Ni$_{2}$FeGa}.
\end{center}
\end{figure}

The  calculations of J$_{ij}$ are performed within a cluster of radius 3$a$, where $a$ is the optimized  lattice parameter of the respective unit cell. To start with we present the Heisenberg coupling parameters for the Ni-based systems in Figure 4. This figure suggests that for all the Ni-based materials the X-Y (X = Ni; Y = Cr, Mn, Fe) interaction is purely ferromagnetic in nature. As X and Y atoms are nearest neighbours, direct interaction between these atoms becomes dominating, and it vanishes at a larger distance. For all the cases, Y-Ga interaction is very weak, so we do not discuss this in further analysis of our results. Now we move our focus to the Y-Y interaction: we find that in Ni$_{2}$MnGa, this interaction is  ferromagnetic up to the fifth coordination shell, which is well supported by the literature\cite{prb-achakrabarti-2013} and it is  RKKY type of interaction\cite{pr-kittel-1954} in nature. However, for Ni$_{2}$FeGa, we find that the interaction becomes oscillatory which clearly indicates the presence of a RKKY type of interaction\cite{pr-kittel-1954} in the system. Similar is the case for the Cr atom, as is observed in case of Ni$_{2}$CrGa.

\begin{figure}[ht]
\begin{center}
\includegraphics[width=0.6\columnwidth]{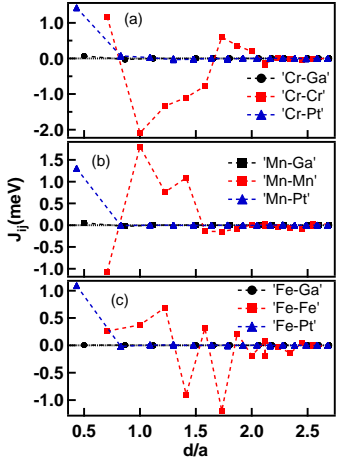}
\caption{Heisenberg exchange parameters J$_{ij}$ of Y atom (Y = Cr, Mn, Fe) with all the atoms as a function of normalized distance $d$$/$$a$ where $a$ is the lattice constant of the austenitic phase for (a)Pt$_{2}$CrGa (b)Pt$_{2}$MnGa and (c)Pt$_{2}$FeGa}
\label{fig:appbimg}
\end{center}
\end{figure}

Figure 5 also suggests that similar to the Ni-based materials, the X-Y (X = Pt; Y = Cr, Mn, Fe) interaction is purely ferromagnetic and it almost vanishes after first nearest neighbour interaction. In case of Pt$_{2}$FeGa, Fe-Fe interaction is very similar to that of Ni$_{2}$FeGa. But for Pt$_{2}$MnGa here first Mn-Mn interaction is strongly anti-ferromagnetic though third, fourth and fifth Mn-Mn interaction is strongly ferromagnetic. From the J$_{ij}$ plot of  Pt$_{2}$CrGa, it is clear that Cr-Cr interaction is more anti-ferromagnetic type, as is observed from Figure 5, though the first neighbour interaction has a ferromagnetic nature.

\begin{figure}[ht]
\begin{center}
\includegraphics[width=0.6\columnwidth]{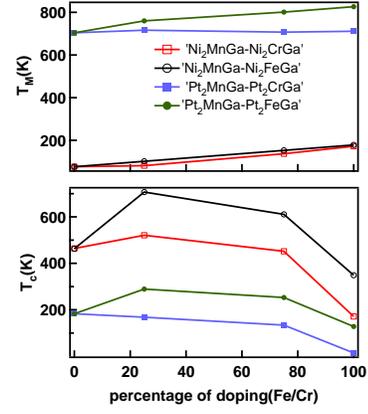}
\caption{Curie temperature(T$_{c}$) and Martensite temperature(T$_{M}$) as a function of substitution}
\label{fig:appbimg}
\end{center}
\end{figure}

In Figure 6, we plot the T$_{M}$ values from our previous calculations\cite{jmmm-troy-2016}. In Figure 6 we also present the Curie temperature(T$_{c}$) for all the materials in the FM state, calculated from the corresponding J$_{ij}$ values under a mean field approximation following the approach of Meinert\textit{ et al}\cite{jpcm-meinert-2011}. Both the temperature values show a systematic change as the composition is changed. We note that the experimental T$_{c}$ of Ni$_{2}$MnGa is reported as 363K\cite{jalcom-nishihara-2007}. It is already reported in literature that mean field approximation overestimates T$_{c}$\cite{prb-pajda-2001}.  It is observed that Ni$_{2}$CrGa among the Ni-based materials 
and Pt$_{2}$CrGa among the Pt-based materials possess the lowest ferromagnetic transition temperature(T$_{c}$). One possible reason behind this may be the presence of anti-ferromagnetic (AFM) type exchange interaction between Cr-Cr for both of them. 

\subsubsection{Anti-ferromagnetic Interaction in Cr-based systems}
\begin{figure}[h]
\begin{center}
\includegraphics[width=0.6\columnwidth]{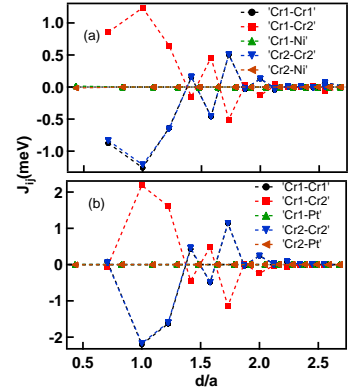}
\caption{Heisenberg exchange parameters J$_{ij}$ of Cr atom  with its neighbours as a function of normalized distance d$/$a where a is the lattice constant of the austenitic phase for (a)Ni$_{2}$CrGa (b)Pt$_{2}$CrGa in their correct magnetic ground state (AFM) configuration}
\label{fig:appbimg}
\end{center}
\end{figure}

There is an indication of presence of anti-ferromagnetic interaction in the Cr-based materials as is seen from Figure 4, 5  and 6. It is known that both Cr and Mn have anti-ferromagnetic interaction in their bulk form. From the Bethe-Slater curve, it is clear that Cr has stronger anti-ferromagnetic J$_{ij}$ interaction compared to Mn. We also note that, Sharma \textit{et al}\cite{jpcm-sharma-2010} reported that the doping of Cr at the Mn site of Ni-Mn-In reduces the FM interaction, while doping by Fe leads to increase of FM interaction. All these observations have motivated us to probe the possibility of anti-ferromagnetic ground state in the cases of Ni$_{2}$CrGa and Pt$_{2}$CrGa in our earlier work\cite{jmmm-troy-2016}. In that work, we have predicted that the true ground state magnetic configuration for  Ni$_{2}$CrGa and Pt$_{2}$CrGa is an {\it intra-sublattice} AFM configuration.
\\In Figure 7, we show the variation of J$_{ij}$ in their magnetic ground state (AFM configuration) of  Ni$_{2}$CrGa and Pt$_{2}$CrGa. The calculations for {\it intra-sublattice} AFM configuration have been performed within coherent potential approximation implemented in SPRKKR code, by considering 50\% of the body-centered site(fractional coordinates 0.5, 0.5, 0.5) is occupied by  up spin Cr atom (Cr1) and rest 50\% of the site is occupied by down spin Cr atom (Cr2). From Figure 7, as expected, we can see that Cr1-Cr1 interaction is almost identical with Cr2-Cr2 interaction for the materials. Also the interaction is of RKKY type as is seen from the J$_{ij}$ plots. One more interesting point for both the systems is that the magnetic interaction between Cr1-Cr1 (also Cr2-Cr2) is exactly cancelled by the Cr1-Cr2 (Cr2-Cr1) interaction. These two strong mutually cancelling exchange interactions may have led to the anti-ferromagnetic ground state magnetic configuration for both of these systems.

\section{Conclusion}  
Using first principles density functional theory, we have discussed the effects of Mn replacement by Fe and Cr in case of Pt$_{2}$MnGa as well as Ni$_{2}$MnGa on the respective electronic density of states as well as magnetic interaction. It is clear that all the Ni and Pt-based materials are having a prominent Ni and Pt d-electron derived peaks, respectively, very close to (also below) the Fermi level. We note that it is expected from their respective DOS that tetragonal distortion may lower the energy of these systems compared to their cubic structure, which can be very well explained by band Jahn-Teller distortion mechanism.  Further, while  for Fe and Mn cases, the minority spin makes the major contribution, in case of Cr, both the spin densities of states show similar intensity. The total and partial spin-polarized DOS as well as the valence band width show similarity between the Pt and Ni-based alloys. 
\\The plots of J$_{ij}$ as a function of normalized inter neighbor spacing suggest that in all the materials X-Y interaction (X = Pt, Ni; Y = Cr, Mn, Fe) has a  direct ferromagnetic interaction. But Y-Y (Y = Cr, Mn, Fe) interaction is mainly RKKY type for both  Fe-based systems i.e. Ni$_{2}$FeGa and Pt$_{2}$FeGa. For Ni$_{2}$MnGa also, the Mn-Mn interaction is RKKY type, with a ferromagnetic interaction up-to the fifth nearest neighbour. On the contrary, in  Pt$_{2}$MnGa the anti-ferromagntic interaction becomes somewhat dominating between Mn-Mn, though we find an overall ferromagnetic state is energetically lower compared to the anti-ferromagnetic state. But in case of Cr substitution at Mn site, in both the cases  Ni$_{2}$CrGa and Pt$_{2}$CrGa, Cr-Cr anti-ferromanetic exchange interaction becomes predominant over the ferromagnetic one, which leads to an {\it intra-sublattice} anti-ferromagnetic configuration in their respective ground state\cite{jmmm-troy-2016} and the  J$_{ij}$ plots for the AFM phase show interesting difference with the corresponding plots of the ferromagnetic configuration of the Cr-based alloys.
\section{Acknowledgement}
Authors thank P. D. Gupta and P. A. Naik for facilities 
and encouragement throughout the work. The scientific computing group,
 computer centre of RRCAT, Indore and P. Thander are thanked for 
help in installing and support in running the codes. S. R. Barman and C. Kamal are thanked for 
useful discussions. TR thanks HBNI, RRCAT for financial support.

\end{document}